\documentclass{article}

\PassOptionsToPackage{numbers, compress}{natbib}
\usepackage[preprint]{neurips_2023}
\usepackage{graphicx}
\usepackage{amsmath}
\usepackage{multirow}
\usepackage[ruled,vlined]{algorithm2e}

\usepackage[utf8]{inputenc} 
\usepackage[T1]{fontenc}    
\usepackage{hyperref}       
\usepackage{url}            
\usepackage{booktabs}       
\usepackage{amsfonts}       
\usepackage{nicefrac}       
\usepackage{microtype}      
\usepackage{xcolor}         
\usepackage{pifont}

\title{Assessing the Extrapolation Capability of \\Template-Free Retrosynthesis Models}

\author{%
  Shuan Chen \qquad Yousung Jung \\
  Department of Chemical and Biological Engineering \\
  Seoul National University \\
  Seoul, South Korea 08826 \\
  \texttt{yousung.jung@snu.ac.kr} \\
}

\begin{document}

\maketitle

\begin{abstract}
Despite the acknowledged capability of template-free models in exploring unseen reaction spaces compared to template-based models for retrosynthesis prediction, their ability to venture beyond established boundaries remains relatively uncharted. In this study, we empirically assess the extrapolation capability of state-of-the-art template-free models by meticulously assembling an extensive set of out-of-distribution (OOD) reactions. Our findings demonstrate that while template-free models exhibit potential in predicting precursors with novel synthesis rules, their top-10 exact-match accuracy in OOD reactions is strikingly modest (\(<1\%\)). Furthermore, despite the capability of generating novel reactions, our investigation highlights a recurring issue where more than half of the novel reactions predicted by template-free models are chemically implausible. Consequently, we advocate for the future development of template-free models that integrate considerations of chemical feasibility when navigating unexplored regions of reaction space.
\end{abstract}

\section{Introduction}
In recent years, the field of scientific research has witnessed remarkable progress in the development of deep learning models. These cutting-edge deep learning models have been increasingly applied to tackle complex scientific challenges, including synthesis planning \cite{schwaller2022machine, tu2023predictive}. Retrosynthesis models relying on pre-defined reaction templates derived from existing reactions are called template-based models. Due to the nature of using known reaction templates, their incapability of predicting novel synthesis pathways is often considered as a major disadvantage \cite{coley2017computer, segler2017neural, dai2019retrosynthesis, chen2021deep}. In contrast, template-free, especially language-based, models, are often claimed to hold the advantage of discovering entirely new reaction rules that have not been encountered during training \cite{liu2017retrosynthetic, tetko2020state, seo2021gta, irwin2022chemformer, wan2022retroformer, tu2022permutation}.

However, the freedom to explore beyond established boundaries in chemical reaction prediction also introduces a potential risk—the generation of invalid or chemically infeasible reactions. While the benefits of template-free models are widely extolled, the associated risks are often underestimated and underexplored. Additionally, it remains unclear whether template-free models can effectively predict out-of-distribution (OOD) synthesis records, further emphasizing the need for a comprehensive examination of their capabilities. In this paper, we evaluate the extrapolation capabilities of template-free models by asking two questions:

\begin{enumerate}
\item Can template-free models accurately reproduce the synthesis pathways recorded in the existing datasets involving unseen reaction templates from the train set?
\item How "novel" are the recently proposed template-free models, and what is the chemical feasibility of the predicted novel reactions? 
\end{enumerate}

We achieve this by meticulously curating an OOD dataset sourced from two different datasets extracted from USPTO reactions \cite{lowe2012extraction}. Additionally, we delve into the examination of the popular novel reaction templates proposed by these models, seeking to gain deeper insights into the practical potential of employing template-free, in particular language-based, models for the discovery of innovative synthesis strategies. Through this research, we aim to provide a nuanced understanding of the capabilities, limitations, and practical implications of template-free models in the context of retrosynthesis prediction. 

\section{Related Work}

\textbf{Template-based models} Template-based models classify which of the predefined reaction templates should be applied to a given product in order to derive its precursors. These templates can be either manually crafted or automatically extracted from extensive reaction databases \cite{hartenfeller2011collection, coley2019rdchiral}. The application of these templates specifies which part of the molecule needs to be modified, facilitating the prediction of precursor molecules based on input molecules \cite{coley2017computer, segler2017neural, dai2019retrosynthesis, chen2021deep}. These models, however, are inherently limited by the finite set of templates they operate within.

\textbf{Template-free models} In contrast, template-free models have emerged as a novel paradigm, offering a departure from the constraints of predefined templates. These models predict precursor molecules by directly generating reactants from scratch, without relying on predefined reaction templates. This liberates them from the confines of a predefined template space, allowing for a more flexible and unconstrained approach to retrosynthesis prediction \cite{liu2017retrosynthetic, tetko2020state, seo2021gta, irwin2022chemformer, wan2022retroformer, tu2022permutation}. These models hold the potential to transcend the limitations of predefined templates, enabling them to explore a wider and more diverse reaction space in retrosynthetic prediction.

\section{Defining out-of-distribution reactions}
\label{sec:ood}
To rigorously assess the extrapolation capabilities of template-free retrosynthesis models, we must assemble a set of reactions that distinctly deviate from those present in the training dataset. Our approach commences with two widely recognized reaction datasets for retrosynthesis and forward synthesis model evaluation: the USPTO-50k dataset \cite{schneider2016s} and the USPTO-480k dataset \cite{jin2017predicting}.

Given that the majority of retrosynthesis models are trained on the training set of the USPTO-50k dataset, comprising 40,000 reactions, we derive a set of reaction templates from this training data to represent the learned reaction distribution. The representation of reaction templates in this study employs the concept of Generalized Reaction Templates (GRTs) \cite{chen2022generalized}, chosen for their effective characterization of the dataset's chemical reaction distribution. GRTs, achieved by simplifying reaction templates to remove atom symbols from local reaction template \cite{chen2021deep}, offer a high level of coverage (99.8\%) on the USPTO-50k test set.

Subsequently, we extract reaction templates from the test set of the USPTO-480k dataset, which also encompasses 40,000 reactions, and identify reactions that do not share any GRTs with those seen in the USPTO-50k training set. This subset of reactions is aptly denoted as "out-of-distribution (OOD)" reactions, signifying their divergence from the reaction distribution in the training set. To facilitate comparison, we randomly sample an equivalent number of reactions from the remaining reactions in the USPTO-480k dataset and classify them as "in-distribution (ID)" reactions.

\begin{figure}[ht]
\begin{center}
\includegraphics[width=13.5cm]{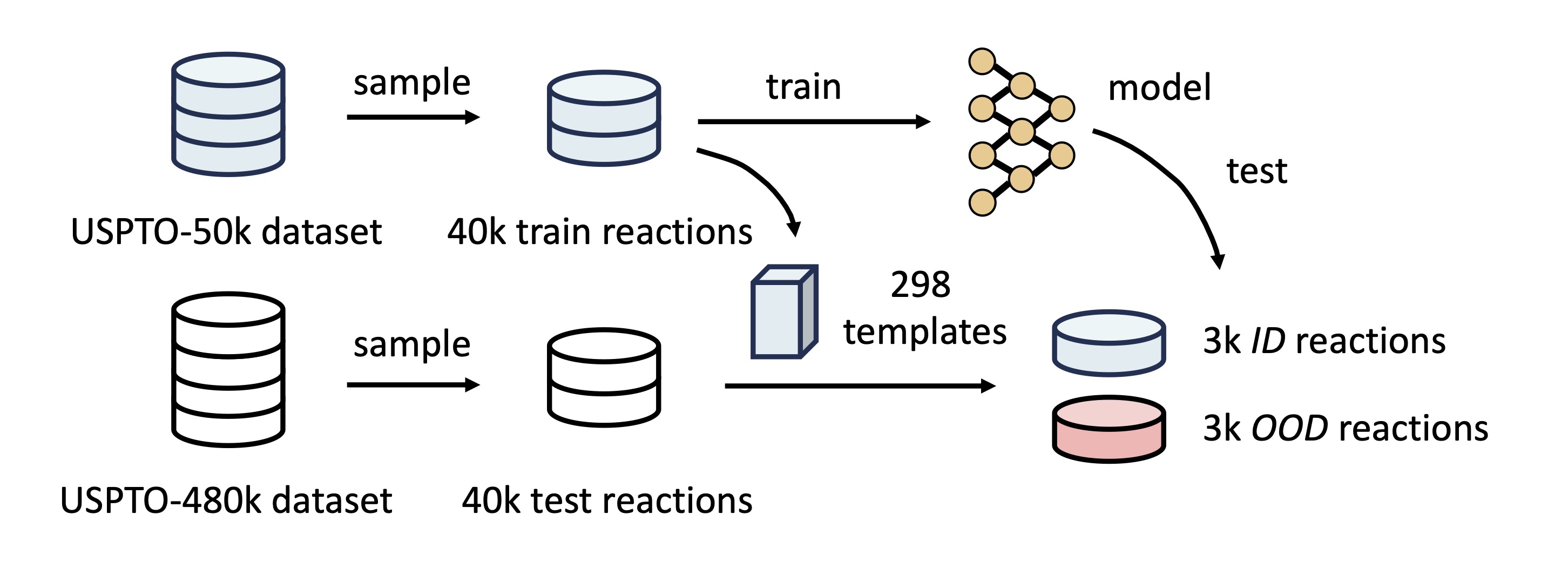}
\end{center}
\caption{The workflow of collecting in-distribution (ID) and out-of-distribution (OOD) reactions}
\label{fig:OOD collection}
\end{figure}

The objective is to assess whether the accuracy of template-free models in predicting reactions from the OOD dataset is on par with their performance on the ID dataset. If template-free models can predict OOD reactions with comparable accuracy to ID reactions, it would suggest their proficiency in extrapolating the reaction space and unveiling novel synthesis rules. The overall process of curating OOD reactions is illustrated in Figure \ref{fig:OOD collection}.

\section{Experiments}
\textbf{Data:} We evaluate the models on both 3,277 in-distribution (ID, reactions involving known reaction templates) and 3,277 out-of-distribution (OOD, reactions not involving known reaction templates) reactions described in Section \ref{sec:ood}.

\textbf{Evaluation metric:} We use (1) exact-match accuracy, (2) round-trip accuracy, (3) reaction validity, (4) reaction uniqueness, and (5) reaction novelty as described below to evaluate the predictions:

\textit{Exact-match accuracy:} Exact-match accuracy is determined by whether the canonical SMILES strings of predicted reactants matches the ground truth reactants in the dataset.

\textit{Round-trip accuracy:} To assess the chemical feasibility of reactions predicted by template-free models, we employ the Molecular Transformer \cite{schwaller2019molecular} to predict the outcomes of the proposed precursors. A successful match between the predicted reactant outcomes (top-1 prediction) and the original target product confirms the validity of retrosynthesis predictions, referred to as "round-trip accuracy"\cite{schwaller2020predicting}.

\textit{Reaction validity:} The evaluation of prediction validity entails two criteria: (1) Predicted SMILES strings can be successfully converted into molecules using RDKit \cite{landrum2013rdkit}. (2) Predicted reactants include all the required atoms to form the product, ensuring a balanced reaction.

\textit{Reaction uniqueness:} The uniqueness of predictions is quantified by calculating the ratio of non-repeated canonical SMILES strings among all the generated predictions.

\textit{Reaction novelty:} To assess the novelty of predictions, we analyze whether the reactions yield templates that have never appeared in the USPTO-50k training set. Atom-to-atom mappings for template extraction are generated using RXNMapper \cite{schwaller2021extraction} to facilitate this evaluation.

\textbf{Models:} We implement three template-free models, including the vanilla Transformer \cite{vaswani2017attention}(trained from scratch using USPTO-50K train set, referred to as Transformer in this paper), and its two variants, Retroformer \cite{wan2022retroformer} and Chemformer \cite{irwin2022chemformer}. For comparison, we compare the prediction results with a template-based baseline model LocalRetro \cite{chen2021deep}.

Note that although Zhong et al. \cite{zhong2022root} reported higher accuracy on the USPTO datasets, we found a potential data leakage issue arising from pretraining on USPTO molecules. Similarly, $O$-GNN \cite{zhu2022mathcal}, a variant of LocalRetro, reported higher accuracy than LocalRetro. However, upon closer examination, we discovered that its marginal improvement is more likely attributed to hyperparameter tuning rather than fundamental advancements in the model design. Consequently, these two models are not included in this paper for comparison.

\subsection{Top-k exact-match and round-trip accuracy}
Table \ref{table:exact-match} presents the top-k exact-match accuracy of four tested retrosynthesis models, including one template-based and three template-free models, on in-distribution (ID) and out-of-distribution (OOD) reactions. In the case of ID reactions, all template-free models demonstrate lower accuracy than the template-based model LocalRetro, with LocalRetro achieving the highest performance, particularly outperforming by over 15.7\% at \textit{k}=10. Conversely, for OOD reactions, template-free models successfully predict synthesis steps using reaction templates not seen in the USPTO-50k train set, while LocalRetro, by definition, fails to predict any of them. However, even at top-10 predictions, Transformer, Retroformer, and Chemformer only correctly predict 9 (0.27\%), 26 (0.79\%), and 9 (0.27\%) out of the total 3,277 reactions, which is significantly lower than their performance on ID reactions (78.9\%, 79.0\%, and 70.3\%, respectively).

\begin{table}[ht]
\caption{Top-k exact-match accuracy (\%) on in-distribution (ID) and out-of-distribution (OOD) reactions of four retrosynthesis models. The highest values are highlighted in bold font.}
\begin{center}
\begin{tabular}{lccccccc}
 \hline
 \multirow{2}{*}{Model}& \multicolumn{3}{c}{ID reactions} && \multicolumn{3}{c}{OOD reactions}\\
 \cline{2-4} \cline{6-8}
  & \textit{k}=1&5&10 && \textit{k}=1&5&10\\
 \hline
 LocalRetro \cite{chen2021deep}  &\textbf{64.3} &\textbf{89.8} &\textbf{94.7} &&0 &0 &0\\
 \hline
 Transformer \cite{vaswani2017attention} &52.6 &75.0 &78.9 &&0.09 &0.24 &0.27\\
 Retroformer \cite{wan2022retroformer} &59.9 &72.9 &79.0 &&0.03 &\textbf{0.31} &\textbf{0.79}\\
 Chemformer \cite{irwin2022chemformer} &63.6 &70.0 &70.3 &&\textbf{0.24} &0.27 &0.27\\
 \hline
\end{tabular}
\end{center}
\label{table:exact-match}
\end{table}

Table \ref{table:round-trip} illustrates the top-10 round-trip accuracy of three template-free models. The round-trip accuracy shares a similar trend across two test sets. For predicted reactions corresponding to known reaction templates, Chemformer demonstrates the highest round-trip accuracy. Surprisingly, the vanilla Transformer exhibits nearly double the round-trip accuracy compared to two advanced models.

In general, the round-trip accuracy of predictions yielding novel reaction templates is significantly lower for both Retroformer and Chemformer compared to reactions with known templates. Specifically, Retroformer shows a reduction of 57.4\% and 48.9\% on ID and OOD reactions, respectively. Similarly, Chemformer displays a decrease of 65.5\% and 48.8\% on ID and OOD reactions, respectively. In contrast, the round-trip accuracy of the vanilla Transformer remains consistent for predictions involving either known or novel reaction templates.

\begin{table}[ht]
\caption{Top-10 round-trip accuracy (\%) for predictions yielding known templates (\textit{known}) and novel templates (\textit{novel}) on ID and OOD reactions. The highest values are highlighted in bold font.}
\begin{center}
\begin{tabular}{lccccc}
 \hline
 \multirow{2}{*}{Model}& \multicolumn{2}{c}{ID reactions} && \multicolumn{2}{c}{OOD reactions}\\
 \cline{2-3} \cline{5-6}
  & \textit{known} & \textit{novel} && \textit{known} & \textit{novel}\\
 \hline
 Transformer \cite{vaswani2017attention} &54.1 &\textbf{55.9} &&43.3 &\textbf{42.5}\\
 Retroformer \cite{wan2022retroformer} &84.2 &26.8 &&70.4 &21.5\\
 Chemformer \cite{irwin2022chemformer} &\textbf{88.4} &22.9 &&\textbf{73.2} &24.4\\
 \hline
\end{tabular}
\end{center}
\label{table:round-trip}
\end{table}

Note that the top-1 accuracy of the pretrained Molecular Transformer\cite{schwaller2019molecular}, which was used to evaluate the round-trip accuracy and trained on USPTO-480k dataset \cite{jin2017predicting}, on the tested reactions is 94.0\%.

\subsection{Validity, uniqueness and novelty}

Table \ref{table:quality} provides an analysis of the validity, uniqueness and novelty of the top-10 predictions from three template-free models on ID and OOD reactions. In the case of ID reactions, Chemformer demonstrates a higher rate of valid predictions (98.6 \%) but significantly lower uniqueness (15.6 \%) and novelty (1.6 \%) compared to Transformer and Retroformer. The low number of unique reactions predicted by Chemformer may contribute to its higher top-1 accuracy but lower accuracy at higher \textit{k} compared to other non-pretrained models. Although Transformer and Retroformer generates much more novel reactions than Chemformer, the accuracy of top-10 predictions on OOD reactions does not scale similarly, indicating the predicted novel reactions are mostly different from the recorded reactions. Interestingly, all template-free models generate more unique and novel reactions when predicting retrosynthesis for the targets of OOD reactions than for of ID reactions, indicating a potential understanding by template-free models of the need to extrapolate the reaction space learned from the training set to make correct OOD predictions.

\begin{table}[ht]
\caption{The percentage of valid, unique, and novel reactions tested on ID and OOD reactions within top-10 predictions from template-free models. The highest values are highlighted in bold font.}
\begin{center}
\begin{tabular}{lcccccccccc}
 \hline
  \multirow{2}{*}{Model}& \multicolumn{3}{c}{ID reactions} && \multicolumn{3}{c}{OOD reactions}\\
 \cline{2-4} \cline{6-8}
  & \textit{\% Valid} & \textit{\% Unique} & \textit{\% Novel} && \textit{\% Valid} & \textit{\% Unique}& \textit{\% Novel}\\
 \hline
 Transformer \cite{vaswani2017attention} &55.3 &49.5 &\textbf{14.7} &&58.6 &54.4 &17.6\\
 Retroformer \cite{wan2022retroformer} &81.0 &\textbf{60.7} &12.6 &&76.3 &\textbf{67.0} &\textbf{18.2}\\
 Chemformer \cite{irwin2022chemformer} &\textbf{98.6} &15.6 &1.4 &&\textbf{96.6} & 21.9 & 1.6\\
 \hline
\end{tabular}
\end{center}
\label{table:quality}
\end{table}

\subsection{Investigating novel reactions}
\begin{figure}[ht]
\begin{center}
\includegraphics[width=13.5cm]{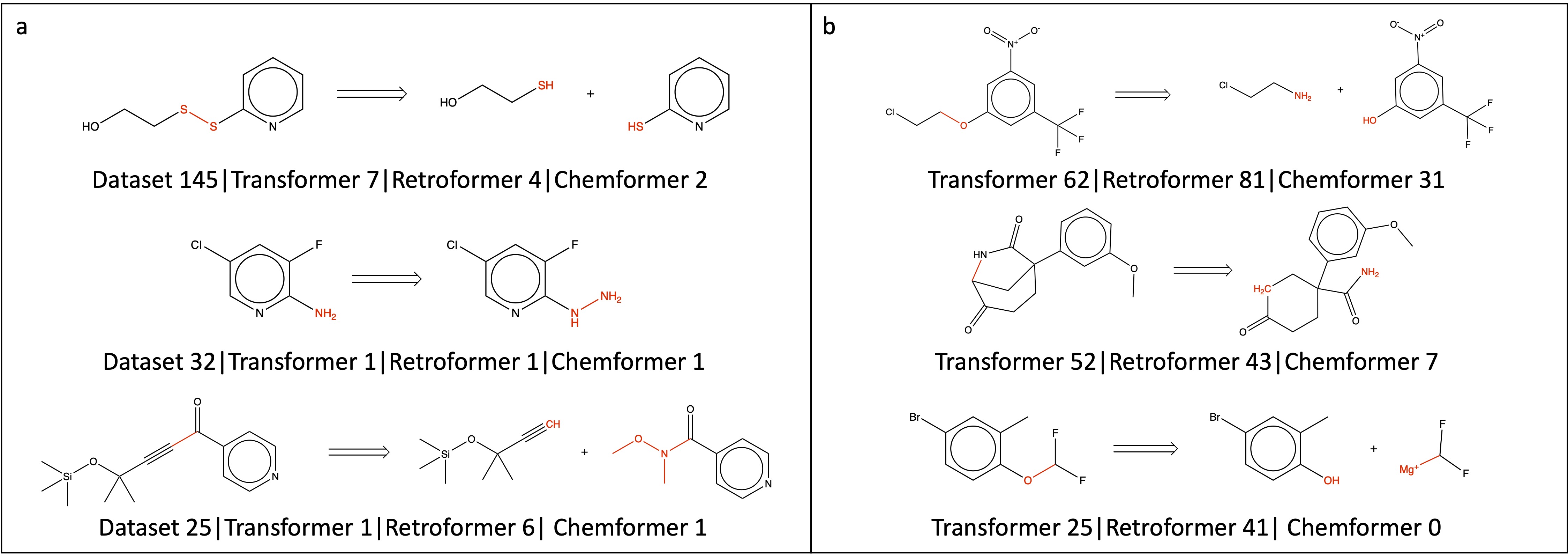}
\end{center}
\caption{Examples of (a) correctly predicted and (b) popular novel reaction templates generated by Transformer, Retroformer, and Chemformer. The number of reactions corresponding to each reaction template in the dataset or predicted by each model is written below each example reaction.}
\label{fig:novel-templates}
\end{figure}

In this section, we delve into the analysis of novel reaction templates generated by three template-free models to gain a deeper understanding of the extrapolation achieved within their training data. In Figure \ref{fig:novel-templates}a, we present three examples of out-of-distribution (OOD) reactions that were correctly predicted by all template-free models using different reaction templates.  While these template-free models demonstrate the ability to generate chemically reasonable templates for these examples, the accuracy in predicting such templates remains relatively low, standing at 4.8\%, 3\%, and 4\% for Transformer, 2.8\%, 3.1\%, and 24\% for Retroformer, and 1.4\%, 3\%, and 4\% for Chemformer for each of the template examples showcased in Figure \ref{fig:novel-templates}a.

We also explore the popular novel reaction templates generated by three tested models in Figure \ref{fig:novel-templates}b, aiming to ascertain if template-free models extrapolate further into the reaction space beyond what is recorded in the OOD reactions collected in this study. Our analysis reveals that these so-called "novel templates" often involve the use of nucleophiles as leaving groups (as observed in the amine and Grignard reagent reactions) or direct cleavage of a C-N single bond (as seen in the middle example). While the reaction templates involved in these predictions have not been encountered in the training set, they are likely to be chemically unreasonable. This observation of chemically implausible reactions extends to the remaining novel reactions as well. 

\section{Conclusion}
In summary, this paper has conducted a comprehensive evaluation of the extrapolation capabilities of template-free models in retrosynthesis, specifically focusing on their potential to discover new synthesis strategies beyond the confines of training data. While template-free models may exhibit lower accuracy when predicting in-distribution reactions compared to template-based models, our results affirm their ability to generate reactions documented in the out-of-distribution (OOD) dataset. However, given the extremely low exact-match accuracy on OOD reactions and poor round-trip match accuracy on novel predictions, a substantial advancement of template-free models is necessary before confidently asserting their capability to extrapolate the reaction space. 

Therefore, as we look to the future, the development of template-free models must place a strong emphasis on considering chemical feasibility while navigating and extending the boundaries of the reaction space. This consideration is paramount in ensuring that the innovative potential of template-free models can be harnessed effectively for the advancement of the field of organic chemistry and retrosynthesis.

\bibliographystyle{nips}
\bibliography{neurips_2023}

\appendix

\end{document}